\documentclass[aps,prl,twocolumn,floatfix,superscriptaddress,graphics,longbibliography]{revtex4-1}

\usepackage{soul}
\usepackage{bm}
\usepackage[retainorgcmds]{IEEEtrantools}
\usepackage{graphicx}
\usepackage{mathrsfs}
\usepackage{amsmath}
\usepackage{amssymb}
\usepackage{xcolor,colortbl}
\usepackage{mathrsfs}  
\usepackage{amsfonts}
\usepackage{amsthm}
\usepackage{times,txfonts}
\usepackage{nicefrac}
\usepackage[colorlinks=true,linkcolor=blue,urlcolor=blue,citecolor=blue,pdfusetitle]{hyperref}
\usepackage{framed}

\newcommand{\tr}{\mathrm{tr}}

\newcommand{\change}[1]{ \textcolor{black}{#1}}

\newcommand{\acorr}[2]{{\color{blue}\st{#1}} {\color{red} #2}}

\begin{document}

\title{Machine classification for probe-based quantum thermometry}
\date{\today}
\author{Fabrício S. Luiz}
\affiliation{Faculdade de Ciências, UNESP - Universidade Estadual Paulista, 17033-360 Bauru, S\~ao Paulo, Brazil}
\author{A. de Oliveira Junior}
\affiliation{Faculty of Physics, Astronomy and Applied Computer Science, Jagiellonian University, 30-348 Kraków, Poland.}
\author{Felipe F. Fanchini}
\affiliation{Faculdade de Ciências, UNESP - Universidade Estadual Paulista, 17033-360 Bauru, S\~ao Paulo, Brazil}
\author{Gabriel T. Landi}
\affiliation{Instituto de F\'isica da Universidade de S\~ao Paulo,  05314-970 S\~ao Paulo, Brazil.}

\begin{abstract}
We consider probe-based quantum thermometry and show that machine classification can provide \change{model-independent estimation with quantifiable error assessment.}
Our approach is based on the $k$-nearest-neighbor algorithm.
\change{
The machine is trained using data from either  computer simulations or a calibration experiment.
}
This yields a predictor which can be used to estimate the temperature from new observations.
The algorithm is highly flexible and works with any kind of probe observable.
It also allows to incorporate \change{experimental errors, as well as uncertainties about experimental parameters.}
\change{We illustrate our method with an impurity thermometer in a Bose-gas, as well as in the estimation of the thermal phonon number in the Rabi model.}


\end{abstract}

\maketitle

{\bf \emph{Introduction-}} 
Measuring the temperature of a body has long been a fundamental task in science and technology. 
The enormous range of scales involved, from cosmology to ultra-cold gases, motivate the development for a wide variety of strategies. The drive toward the microscale has been pushing the development of novel methods~\cite{Giazotto2006,Yue2012,Onofrio2017,Karimi2020,Gasparinetti2015}, and recent advances in platforms such as ultra-cold atoms~\cite{Sabin2014,Marzolino2013,Mehboudi2018a,Mitchison2020,Bouton2020}, nitrogen-vacancy centers~\cite{Neumann2013,Shim2021} and superconducting circuits~\cite{Halbertal2016}, have opened up  entirely new frontiers~\cite{Mehboudi2018,Pasquale2018}.

There have been significant advances in understanding the ultimate bounds on thermometric precision, which were analyzed in a variety of models~\cite{Jevtic2015b,Correa2015,DePasquale2016,Johnson2016,Hofer2017a,Mancino2017a,Campbell2017,Correa2017c}.
If the temperature is estimated from direct measurements in the system, the optimal strategy consists of performing projective measurements in the energy eigenbasis~\cite{Correa2015,Paris2016,Campbell2018a}.
Such a strategy, however, is seldom realistic.
Instead, a more tractable scenario is that of probe-based thermometry, where the temperature of a system is estimated by first allowing it to interact with a probe and then measuring the probe. 
Impurities in ultra-cold gases represent a prototypical example~\cite{Sabin2014,Marzolino2013,Mehboudi2018a,Mitchison2020,Bouton2020}, but several experimental platforms also fit this description. 
For instance, the phonon occupation number of a trapped ion~\cite{Wineland1998,Leibfried2003} or a mechanical resonator~\cite{Bowen2016}, are often estimated from quantum optical measurements, and hence use light as the probe.

A single probe may be repeatedly measured~\cite{DePasquale2017}, or multiple probes may be sent sequentially~\cite{Seah2019,Shu2020}. 
In Ref.~\cite{Hovhannisyan2020} it was recently shown that even using  a single-qubit probe one can still retain $\sim 64\%$ of precision (as compared to a direct measurement), provided  optimal strategies are used.
However, these studies focus on precision bounds,
and most existing strategies for building actual estimators are highly model dependent~\cite{Rubio2020,Alves2021}.
For instance, Ref.~\cite{Mitchison2020} analyzed the dephasing factor of impurities in cold Fermi gases. 


\begin{figure}
    \centering
    \includegraphics{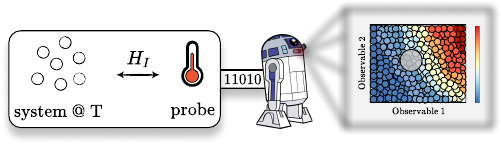}
    \caption{Probe-based thermometry and machine classification. The temperature of a system is estimated by coupling it to a probe, which is subsequently measured. Machine classification uses previously trained data to predict the  temperature from experimental observations. Here we use the KNN algorithm, which constructs an observation heat map (right plot) from training set consisting of pairs $(D_i,T_i)$, corresponding to $d$-dimensional data $D_i$ (here $d = 2$) and associated temperatures $T_i$.  
    }
    \label{fig:diagram}
\end{figure}

In this letter we show how machine classification algorithms can be used to provide precise temperature estimation, in a flexible and experimentally friendly way. 
The scenario we consider is shown in Fig.~\ref{fig:diagram}. 
The temperature $T$ of a system S is measured by first sending a probe P to interact with it, and then measuring the probe. 
This yields some data $D$, from which we want to construct a reliable estimator $\hat{T}(D)$.
Classification accomplishes this by training an algorithm in advance, with a set of points $(D_i,T_i)$.
This can be obtained from, e.g., computer simulations or \change{a calibration} experiment. 
The result is a predictor function, $\hat{T}(D)$, which can be used to estimate the temperature given any real observation $D$. 
Classification \change{is a non-parametric technique}, and hence is model independent, \change{making it extremely flexible}.
It accepts any kind of probe observable, and any kind of S-P interaction strategy. 
Moreover, \change{it is also guaranteed to be asymptotically converge to the true temperature provided the number of training features  increased}~\cite{Samworth2012,Cabannes2021,Duda2000}.


Machine learning has recently seen an explosion of new applications in physics~\cite{DasSarma2019,Carleo2019}, from quantum phase transitions~\cite{Carrasquilla2017,Rem2019,Canabarro2019,Venderley2018} to quantum dynamics~\cite{Vicentini2019,Hartmann2019,Flurin2020,Mazza2021,Martina2021,Fanchini2021,Innocenti2020,Bukov2018,Sgroi2021} and adaptive estimation~\cite{Hentschel2010,*Hentschel2011,Lovett2013,Xu2019b,Schuff2020,Spagnolo2019}.
We will show below that classification in thermometry is \change{robust against many issues commonly faced in realistic thermometry scenarios}.
First, it naturally handles experimental noise. And second, and most remarkably, it handles cases where other parameters in the process are not known. 
For instance, \change{we explore the scenario in which} the S-P interaction strength \change{is only known to lie within a certain range}, which is very reasonable from an experimental point of view. 
\change{Our methods are illustrated  in two experimentally relevant models: impurity thermometry in a Bose-Einstein condensate, and estimation of the thermal phonon number in the Rabi model. }

{\bf \emph{Probe-based thermometry-}} We consider the setting depicted in Fig.~\ref{fig:diagram}. 
A system S,  prepared in a thermal Gibbs state $\rho_S = e^{-\beta H_S}/Z$, at a certain (unknown) inverse temperature $\beta = 1/T$, is coupled to a probe P prepared in an initial state $\rho_P$. 
The total Hamiltonian is  taken as 
$H_{\rm tot} = H_S + H_P + H_I$, where $H_I$ is their interaction. The state of the probe after a certain time $t$ will then be given by $\rho_P(t) = \tr_S \Big\{e^{-i H_{\rm tot}t} \big( \rho_S \otimes \rho_P) e^{i H_{\rm tot} t}\Big\}$, from which information about $T$ can be extracted.

We assume this is accomplished by measuring the expectation values of some probe observables \change{ $\langle \mathcal{O} \rangle_t := \tr\big(\mathcal{O} \rho_P(t)\big)$.
The uncertainty $\delta T^2$ resulting from $\nu$ measurements (obtained in independent repetitions of the experiment) is then~\cite{Braunstein1994,Toth2014,Mehboudi2018a}
\begin{equation}\label{error_bound}
    \delta T^2 = \frac{\Delta^2 \mathcal{O}}{\nu \chi_T^2(\mathcal{O})}, 
\end{equation}
where 
$\Delta^2 O = \langle \mathcal{O}^2\rangle_t - \langle \mathcal{O} \rangle_t^2$ and 
$\chi_T^2(\mathcal{O}) = \partial_T \langle \mathcal{O}\rangle_t$.
Some observables are more sensitive than others; the ultimate precision is determined by the Cramer-Rao bound~\cite{Braunstein1994,Paris2009}
\begin{equation}\label{QFI}
    \delta T^2 \geqslant \frac{1}{\nu \mathcal{F}(T)},
\end{equation}
where $\mathcal{F}(T)$ is the Quantum Fisher Information (QFI). 
When the probe fully thermalizes with the system, the QFI can be written solely in terms of the probe's energy variance~\cite{Pasquale2018,Mehboudi2018}. But in general\acorr{}{,} the state of the probe is out of equilibrium and the QFI must be determined with the usual quantum metrology tools~\cite{Paris2009}. 
}

\change{
Classification can make use of not only a single observable, but a \emph{dataset} $D = (\langle \mathcal{O}_1 \rangle, \ldots, \langle \mathcal{O}_{d}\rangle)$, of dimension $d$. 
This could mean different observables, or the same observable measured at different times. 
In either case, each observable is determined from  independent experiments.
}
Intuitively speaking, the richer the dataset, the less likely it is that the data was generated from any other temperature than the real one. 

{\bf \emph{The $k$ nearest-neighbors (KNN) algorithm -}}
We introduce the KNN classification algorithm~\cite{International2016,Altman1992,Duda2000} as a model-independent (non-parametric) approach to thermometry. 
Classification is a pattern recognition method~\footnote{Other methods, such as neural networks, were also tested, but presented no visible advantages.}. 
We first train the algorithm using $N$ datasets $(D_i,T_i)$ generated from either computer simulations, or a calibration experiment. 
Each dataset $D_i$ is pictured as a point in a $d$-dimensional grid (Fig.~\ref{fig:diagram}), which is also labeled by the corresponding temperature $T_i$. 
When an actual observation $D$ arrives, the algorithm locates its position in this grid and computes the Euclidean distance to its $k$ nearest-neighbors. 
\change{The inverse distances serve as weights to build the probability that $D$ is associated with each $k$ neighbor. 
The average of said probability yields the estimator $\hat{T}(D)$. 
And the variance yields the so-called excess risk $\delta T_{\rm exc}^2$,  which represents the additional uncertainty 
incurred from using a finite number $N$ of training points (which vanishes if $N \to \infty$). 
From $\delta T_{\rm exc}$ we can then compute the mean-squared error (MSE), which also takes into account  the bias:
\begin{equation}\label{MSE}
    {\rm MSE}(D,T) = \delta T_{\rm exc}^2 + \Big(\hat{T}(D) - T\Big)^2,
\end{equation}
with $T$ being the real temperature. 
The MSE can only be estimated if the true temperature is known in advance. Hence, although it serves as a useful figure of merit, one generally would not have direct access to it in an experiment.
The KNN algorithm is asymptotically unbiased~\cite{Duda2000,Samworth2012}, so the MSE also vanishes when $N \to \infty$.
}
In the applications below, we have used the KNN implementation in Python from Ref.~\cite{Pedregosa2011}.

\change{
{\bf \emph{Impurity thermometry in a Bose-Einstein condensate (BEC) - }}
To illustrate the main idea, we start with the experimentally meaningful problem of estimating the temperature of a Bose gas by means of an impurity, for which the BEC acts as a bath~\cite{Olf2015,Sabin2014,Mehboudi2018a}.
We follow an approach similar to~\cite{Mehboudi2018a,Khan2021} and consider a Yb impurity (the probe), trapped in a parabolic potential of frequency $\Omega$, and immersed in a K BEC with trap frequency $\omega_B$. 
The solution for the reduced dynamics of the impurity is given in~\cite{Lampo2018}, and is not restricted to weak coupling.
To illustrate the method, we focus on the steady-state fluctuations of the impurity's position, which reads~\cite{Lampo2018}
\begin{equation}\label{bose_x2}
    \langle x^2 \rangle = \frac{\hbar}{2\pi}\int\limits_{-\omega_B}^{\omega_B} d\omega~ \coth(\hbar\omega/k_B T)\tilde{\chi}''(\omega), 
\end{equation}
where $\tilde{\chi}''(\omega) = (\omega\zeta/m_I)/[(\omega\zeta)^2 + (\Omega^2- \omega^2 + \omega \theta)^2]$ is the impurity's response function, with $\zeta = \pi \gamma \omega^3/2\omega_B^3$ and 
$\theta = -(\gamma\omega/2\omega_B^3) [ \omega_B^2 + \omega^2 \ln((\omega_B/\omega)^2-1)]$.  
Here, $m_I$ is the impurity's mass and $\gamma$ is a constant proportional to the BEC-impurity interaction strength (see~\cite{Lampo2018} for the full Hamiltonian).
}

\begin{figure}
    \centering
    \includegraphics[width=0.45\textwidth]{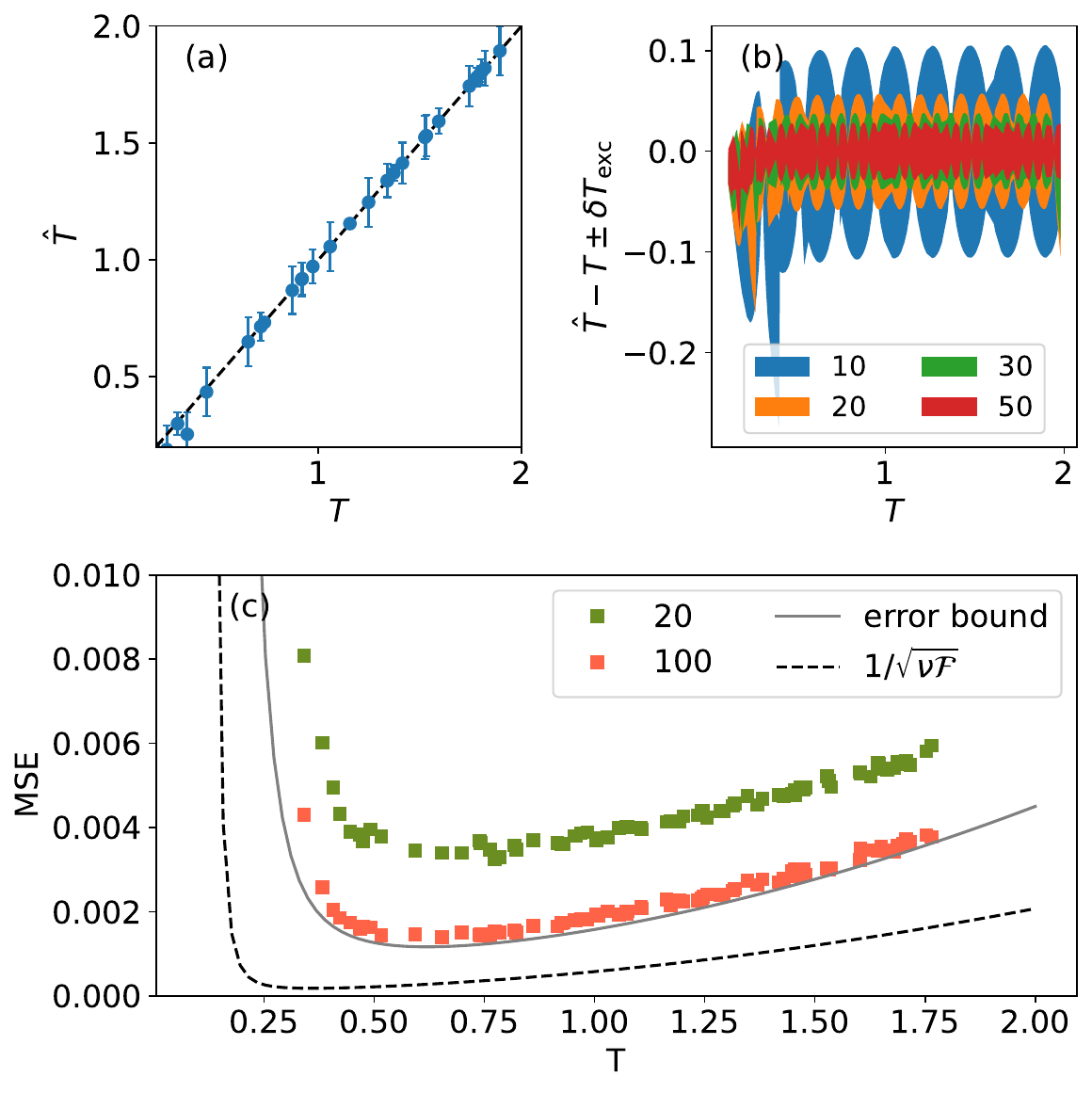}
    \caption{
    \change{
    Impurity thermometry of a BEC, focusing on the steady-state position fluctuations~\eqref{bose_x2}.
    (a) Estimator $\hat{T}$ versus the real temperature $T$ using $N = 10$ points in the training set. Error bars represent the excess risk $\delta T_{\rm exc}$, and can be suppressed by increasing $N$. 
    This is further shown in (b), which plots $\hat{T}-T \pm \delta T_{\rm exc}$ for different choices of $N$. 
    (c) Average MSE from $\nu = 2000$ repetitions of a noisy experiment, for $N = 20$ and $N= 100$. Gray and black-dashed curves corresponds to Eqs.~\eqref{error_bound} and~\eqref{QFI} respectively.
    When only $\langle x^2\rangle$ is measured, no estimator can improve below the gray curve.
    }
    }
    \label{fig:bose}
\end{figure}

\change{
We fix $\gamma = 30$~Hz, $\omega_B = 2\Omega = 2\pi \times 50$~Hz. 
As a first test, we assume that $\langle x^2 \rangle$  can be measured with infinite precision. 
To train the algorithm, we generate pairs $(D_i, T_i)$  with $N$ equally spaced temperatures from $0.1$ nK to $2$ nK.  
The algorithm is then tested using values of  $\langle x^2 \rangle$ obtained from randomly chosen temperatures within the same interval.
Fig.~\ref{fig:bose}(a) shows the predictions $\hat{T}$ as a function of the real temperatures $T$, using only $N = 10$ training points. 
The error bars represent the excess risk $\delta T_{\rm exc}$.
In Fig.~\ref{fig:bose}(b) we plot the difference $\hat{T} - T \pm \delta T_{\rm exc}$ for varying sizes $N$ of the training set.
Small values of $N$  lead to large uncertainties and systematic biases, specially at the boundaries. 
But both are rapidly suppressed with increasing $N$.
}

\change{Next we turn to noisy datasets. In principle, noise could also be included in the training set,  e.g. when the data is obtained from another calibration experiment. 
In the present case, however, the training set is based on the analytical model~\eqref{bose_x2}, and is hence error-free. 
Fig.~\ref{fig:bose}(c) shows the average MSE~\eqref{MSE}, obtained from $\nu = 2000$ independent experiments, for either  $N = 20$ or $N = 100$ training points. 
We also plot Eq.~\eqref{error_bound} in gray, and the Cramer-Rao bound~\eqref{QFI} in dashed (computed  from~\cite{Correa2017c,Mehboudi2018a}).
The latter  can only be reached with special choices of measurement operators~\cite{Paris2009,Escher2011b}, while  Eq.~\eqref{error_bound} represents the best precision attainable using only measurements of $\langle x^2 \rangle$~\cite{Mehboudi2018a}, as in our case. 
}

\change{
When $N = 20$, the MSE is significantly above the gray curve, but for $N = 100$  both the excess risk $\delta T_{\rm exc}$ and the systematic biases are suppressed,  bringing the MSE very close to~\eqref{error_bound}. 
Our method is thus capable of producing quantitatively precise estimates of $T$. 
The only exception is the boundaries of the training set. This happens because the fluctuations generate points $\langle x^2 \rangle$ associated with temperatures outside the interval. 
In real experiments, it is important to avoid this by ensuring the span of the training set is sufficiently broad.
There are also extensions of the KNN algorithm which can monitor whenever a point lies outside the training set, a problem known as anomaly detection~\cite{Gu2019,Bergman2020}.
}

\begin{figure*}
    \centering
    \includegraphics[width=0.9\textwidth]{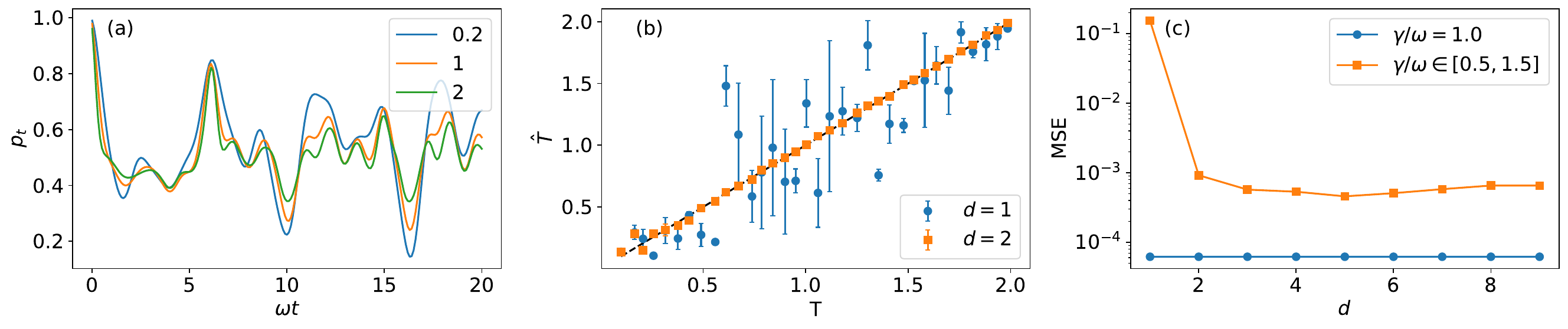}
    \caption{Temperature prediction in the Rabi model. 
    \change{(a) Population $p_t$ for $\gamma/\omega = 1$ and different values of $k_B T/\hbar\omega$. 
    (b) Predicted~vs.~real temperatures for $d = 1, 2$. 
    (c) MSE~vs.~$d$, when $\gamma/\omega$ is known with certainty or when it is only known to lie within a certain interval. 
    The algorithm was trained by generating values of $p_t$, with equally spaced tuples of $(T,\gamma)$ in the intervals $[0.1,2]$ and $[0.5,1.5]$, at times $t = 0.5, 1.0, 1.5, 2.0, \ldots$.
    A dataset with e.g. $d=3$ points consists in the array $D = (p_{0.5}, p_{1.0}, p_{1.5})$.
    }
    }
    \label{fig:Rabi}
\end{figure*}

{\bf \emph{Rabi model-}} 
\change{The previous model served to illustrate how our method can efficiently handle realistic noise in the measurement data. 
But the model itself was far too simple, as it involved only a single feature $\langle x^2 \rangle$, which could also be computed analytically. We now turn to a more complicated model with two new ingredients: (i) the dynamics are not analytically soluble; and (ii) the system-probe interaction strength is not known. 
The latter, in particular, is a very realistic assumption, which is seldom considered in studies of probe-based thermometry. 
Our algorithm can handle this efficiently using additional features ($d > 1$) in the dataset.
This combination of flexibility and robustness is the main advantage of our framework. 
}

We illustrate the idea using the \change{Rabi} model, which frequently appears in a variety of platforms, from cavity quantum electrodynamics to trapped ions and superconducting circuits. 
\change{Similar results can also be obtained, e.g., for the Jaynes-Cummings model.}
The probe is a qubit, with Pauli operators $\sigma_\pm$, and the system is a bosonic mode, with annihilation operators $a$. The total Hamiltonian is
\begin{equation}\label{H_Rabi}
    H = \hbar\omega a^\dagger a + \hbar\Omega  \sigma_+ \sigma_- + \hbar\gamma (a + a^\dagger)(\sigma_+ + \sigma_-),
\end{equation}
where $\gamma$ is the interaction strength.
\change{Estimation of the thermal occupation number of the bosonic mode is one of the most basic problems in e.g., trapped ions~\cite{Leibfried2003,Wineland1998}.
}
Quantities are measured in units of $\omega=1$.
The probe is taken to be resonant with the system ($\Omega = \omega$) and start in the
\change{excited state $\rho_P = |1\rangle \langle 1|$}.
The free parameters are thus the coupling strength $\gamma$, and the system's initial temperature $T$. 
\change{We focus on the probe's populations $p_t = \langle \sigma_+ \sigma_-\rangle_t$, but the algorithm also works with coherences. 
Numerically simulated curves of $p_t$~vs.~$\omega t$, for different $T$, are shown in Fig.~\ref{fig:Rabi}(a) (c.f.~\cite{Lv2017} for experimental results).
They serve to illustrate the non-trivial temperature dependence, which would be difficult to fit with standard methods (specially taking into account the computational complexity of simulating the model). 
}

\change{We consider $k_B T/\hbar \omega \in [0.1,2]$, and assume $\gamma/\omega$ is only known to lie in the interval $[0.5,1.5]$.
Populations were computed numerically for a grid of $100\times100$ tuples $(T,\gamma)$, and for different times $\omega t = 0.5, 1.0, 1.5, 2.0, \ldots$ (other choices of times only marginally affect the results).  
To analyze the role of  the number of features $d$, we adopt the strategy that a dataset with, e.g. $d=3$ consists of $D = (p_{0.5}, p_{1}, p_{1.5})$, and so on. 
For simplicity, we also assume all data points are noiseless, as the effects of such noise have already been explored in Fig.~\ref{fig:bose}(c).
}

\change{
Fig.~\ref{fig:Rabi}(b) shows the results of the estimation when $d = 1$ and $d = 2$. Since $\gamma$ is not known, using only $d = 1$  yields terrible results. But, remarkably, with as little as $d = 2$ features, the results are already remarkably good. 
We explore this further in Fig.~\ref{fig:Rabi}(c), where the MSE is found to decrease dramatically with increasing $d$ (note the log scale), until saturating at a value that is is ultimately determined by the number of points $N$ in the grid. 
We also show in Fig.~\ref{fig:Rabi}(c) the results which would be obtained if $\gamma$ was known with certainty. In this case, the MSE is independent of $d$, with a value once again determined solely by $N$.
}
Thus, with sufficiently many measurements, the precision becomes roughly independent of our uncertainty in the interaction strength.


{\bf \emph{Thermometric data structures-}} 
The results just presented indicate that the use of classification --- and the KNN algorithm ---  in probe-based thermometry is not only versatile, but also robust.
Similar tests have also been performed in various other systems, such as qudit models and spin chains. And we have also explored a large variety of parameter choices: e.g., resonant~vs.~non-resonant energy gaps in Eq.~\eqref{H_Rabi}, different initial probe states, and so on. Even though the fine details differ from one case to the other, the overall performance is similar in all cases: precise estimation with asymptotically diminishing errors. 

We argue that this happens because the probe observables depend smoothly on $T$. Even though the probe is intrinsically out of equilibrium, the spirit is similar to equilibrium quantities, such as energy, entropy or specific heat.
It is rare, for instance, to find observables that are oscillatory in $T$, or behave very erratically. Instead, this smooth dependence causes the data structures to be segmented into well-defined regions, which is crucial for the KNN performance. Thermometry thus represents a niche within the realm of parameter estimation, where classification methods could prove to be particularly useful.

\begin{figure}
    \centering
    \includegraphics[width=0.5\textwidth]{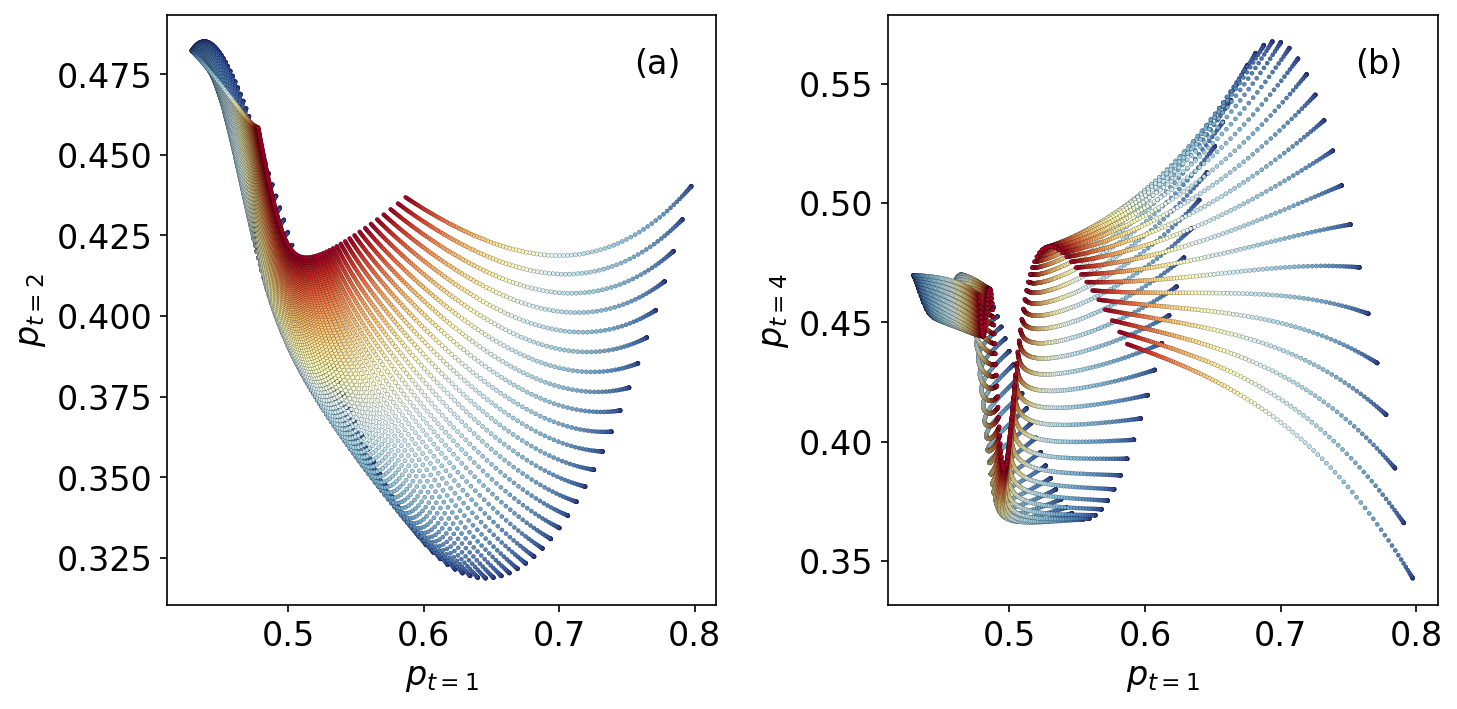}
    \caption{
    \change{Qubit populations in the Rabi model, $p_{t_1}$~vs.~$p_{t_2}$, for different choices of $(T,\gamma)$, with the color of each point representing the corresponding temperature.
    (a) $p_{t=2}$~vs.~$p_{t=1}$.
    (b) $p_{t=4}$~vs.~$p_{t=1}$.
    }
    }
    \label{fig:data_structures}
\end{figure}

To corroborate this argument, we analyze the data structures stemming from the Rabi model~\eqref{H_Rabi}.
\change{Fig.~\ref{fig:data_structures} shows curves of $p_{t_1}$~vs.~$p_{t_2}$ for two choices of $(t_1,t_2)$.
}
The conditions are similar to those of Figs.~\ref{fig:Rabi}.
As can be seen, irrespective of the value of $\gamma$, points are clearly segmented by temperature, and changes from the hot to the cold regions are always smooth.
There are very few regions, for instance, where hot and cold points mix together. 
This explains why the KNN algorithm is successful.    
One should also bear in mind that one often uses more than $d = 2$ observations, \change{which help to further disentangle the cold and hot regions.}

{\bf\emph{Significance- }}
We have showed that classification provides a general and flexible platform, that can be applied to any probe-based system. 
It can accept any kind of observation as input,  handles noise in the dataset, and allows the inclusion of additional uncertainties about the model parameters. 
Moreover, as we have shown, 
\change{it provides quantitative error assessment and is asymptotically consistent.} In light of these facts, we,  believe classification may become a useful tool in experimental quantum thermometry. Indeed, several quantum coherent experiments, such as trapped ions and optomechanics, already fall under this category and could directly benefit from this formalism. 


{\bf\emph{Acknowledgements- }} FFF and GTL acknowledge the financial support of the S\~ao Paulo Funding Agency FAPESP (Grants No.~2021/04655-8, ~2017/50304-7, 2017/07973-5 and 2018/12813-0) and the Brazilian funding agency CNPq (Grant No. INCT-IQ 246569/2014-0). GTL acknowledges the support from the Eichenwald foundation (Grant No.~0118 999 881 999 119 7253). AOJ acknowledges the financial support from the Foundation for Polish Science through TEAM-NET project (contract no.~POIR.04.04.00-00-17C1/18-00). FSL acknowledges financial support of the National Council for Scientific and Technological Development (CNPq) (Grant No. 151435/2020-0).

\bibliography{library}
\end{document}